\documentclass[twocolumn,nofootinbib,prl]{revtex4}

\usepackage{graphicx}
\usepackage{amssymb,amsmath,amsfonts}

\def\beal{{\begin{align}}}
\def\eeal{{\end{align}}}
\def\beq{\begin{eqnarray}}
\def\eeq{\end{eqnarray}}
\def\d{\,{\rm d}}

\def\i{{\rm i}}

\def\thf{\tfrac{1}{2}}
\def\hf{\frac{1}{2}}

\begin{document}

\title{Solution to the problem of time}
\author{Benjamin Shlaer}
\email[ ]{shlaer@cosmos.phy.tufts.edu}

\affiliation{Institute of Cosmology, Department of Physics and Astronomy,\\Tufts University, Medford, MA  02155, USA}
\affiliation{Volen Center for Complex Systems,\\Brandeis University, Waltham, MA  02454, USA}

\begin{abstract}\noindent
Despite the ultraviolet problems with canonical quantum gravity, as an effective field theory its
infrared phenomena
%, including  eternal inflation and black hole evaporation, 
should enjoy fully quantum mechanical unitary time evolution.  Currently this is not possible, the impediment being what is known as the problem of time. 
Here, we provide a solution by promoting the cosmological constant $\Lambda$ to a Lagrange multiplier constraining the metric volume element to
be manifestly a total derivative.  Because $\Lambda$ appears linearly in the Hamiltonian constraint, it unitarily generates time evolution,
yielding a functional Schr\"odinger equation for gravity.   Two pleasant side effects
of this construction are that vacuum energy is dissociated from the cosmological constant problem, much like in unimodular gravity, and the natural foliation
provided by the time variable defines a sensible solution to the measure problem of eternal inflation.    
\end{abstract}

%\pacs{}

\maketitle

The problem of time \cite{Kuchar:1991qf, Isham:1992ms, Anderson:2010xm}, also known as the Hilbert space problem, is the absence of a positive definite probability current that is conserved under time evolution.  It afflicts canonically quantized general relativity,
and is unrelated to the ultraviolet problem of nonrenormalizability, because it also occurs in lower dimensions, where general relativity is renormalizable.
It is often thought that the problem of time arises because of diffeomorphism invariance.  This is not precisely the case, because diffeomorphism invariance is coordinate invariance, and any
theory can be written in coordinate invariant language \cite{Dirac:1967}.   The origin of the problem of time is that in general relativity physical time is a {\em foliation}, essentially determined by the lapse function. 
Since no time derivatives are taken of the lapse function in the gravitational action, its canonical momentum vanishes.  This is a so-called first-class primary constraint, and it ensures that physical-time translation must be arbitrarily chosen in different spatial regions as the spatial geometry is evolved.    
(Note that this is very different from the gauge freedom of coordinate invariance i.e., diffeomorphism invariance, which has no physical content.)
  
In classical general relativity, because the foliation is not determined by the equations of motion, it is a {\em gauge choice}, but this does not imply that it has no coordinate invariant reality.  The foliation for geodesic slicing is not a coordinate transformation from the foliation corresponding to maximal slicing:  They are different physical slicings, having different curvature invariants.  They are only gauge equivalent from the point of view of the resulting 4-geometry that ensues.
Quantum mechanically, all first-class constraints are gauge symmetries \cite{Henneaux:1992ig}, and all states which depend on a gauge choice are projected out of the Hilbert space.  Hence, no states in the physical Hilbert space evolve.  Because classical theories should arise as an $\hbar \to 0$ limit of some quantum theory, this is a pathology of the Einstein-Hilbert action that is independent of its UV completion.

The most commonly attempted solutions to the problem of time involve the introduction of matter clocks, which are dynamical fields that (classically) evolve monotonically, and so might play the role of time \cite{Isham:1992ms}.
However, for Hamiltonians bounded below, all momenta occur quadratically in the Hamiltonian,  so backward propagating modes are inevitably produced \cite{Unruh:1989db}.  This leads to negative probabilities.  It is the first-order characteristic of the time-derivative in the Schr\"odinger equation that guarantees positivity of probabilities under time evolution, i.e., unitarity.

Remarkable progress has been achieved with Lagrange multipliers in Einstein aether theories \cite{Jacobson:2000xp} that use pressureless dust as a generator of unitary time evolution \cite{Kuchar:1990vy, Brown:1994py, Giesel:2007wn, Husain:2011tk}.  The primary difference between these models and our model is that here it is the cosmological constant that becomes a choice of initial conditions, rather than the matter density.  A subtle advantage of this built-in landscape is that the natural foliation does not reward expansion, and so may provide a useful solution to the measure problem of eternal inflation \cite{Guth:2007ng}.

%\subsec{A simple example}
A simple example of the problem of time involves a scalar field $q$ in 0+1 dimensions (quantum mechanics), coupled to a dynamical metric, $\d t^2 = g_{\tau\tau}\d\tau^2$.  Defining the lapse function $\eta(\tau)$ via $\eta^2 = g_{\tau\tau}$,
the action is
\beq\label{eq:SGR1+1}
S[\eta,q] = \int \eta\d\tau \left[ \frac{\dot q^2}{2\eta^2} - V(q) \right].
\eeq
The momentum conjugate to $q$ is $p = \dot q/\eta$, and that conjugate to $\eta$ is $p_\eta \approx 0$.  Here $\approx$
denotes a primary constraint, i.e., a momentum relation that does not determine a velocity.  Instead, the classical velocity $\dot\eta(\tau)$ is an arbitrary gauge choice.
The canonical Hamiltonian is
\beq
H = \eta\left[\frac{p^2}{2} + V(q)\right],
\eeq
which generates $\tau$-translations.
Following Dirac \cite{Dirac:1967}, we ensure that the solution to the primary constraint, $p_\eta~=~0$, is maintained under $\tau$-evolution:
\beq
0 = \dot p_\eta = \{p_\eta,H\}_{\rm P.B.} = -\partial H/\partial\eta,
\eeq
and so we demand $H = 0$.
This is known as a secondary constraint, because it involves the equations of motion.  It is this constraint that preserves coordinate independence (diffeomorphism invariance).
Canonical quantization is achieved by promoting the conjugate momenta to operators satisfying the usual commutation relations and then imposing both constraints, $p_\eta \approx 0$, $H = 0$, as the operator equations
\begin{align}
\hat p_\eta \psi(q,\eta,\tau) &= -i\frac{\partial}{\partial\eta}\psi(q,\eta,\tau)=0, \\
\hat H \psi(q,\eta,\tau) &=\eta\left[-\hf\frac{\partial^2}{\partial q^2} + V(q)\right]\psi(q,\eta,\tau)= 0,
\end{align}
on all physical states $\psi(q,\eta,\tau)$.  Then the time dependent Schr\"odinger equation is
\beq
\i \frac{\partial}{\partial\tau} \psi(q) = \hat H \psi(q) = 0,
\eeq
and so no time-evolution occurs.  
Because energy conservation is reduced to conservation of the number zero, $\tau$-translation is a gauge (i.e. unphysical) symmetry of the theory. 

Perhaps it is not surprising that $\tau$-evolution is a gauge symmetry, since $\tau$ is just a coordinate.  What is surprising is that, even though the classical
theory exhibits {\em proper}-time evolution, the quantum theory does not.  The reason is that the physical foliation is a gauge choice, so any quantum states
that depend on it are projected out.  This violates the correspondence principle --- proper time exists classically
but not quantum mechanically.  Quantization removes so many states that no classical interpretation exists even as $\hbar \to 0$.

%\subsec{Sanity check}
Let us reconsider the above quantum mechanical example, now {\em without} gravity, but still using a diffeomorphism invariant formalism.  We will achieve this by rewriting the familiar non-relativistic action using the identity $\d t = \frac{\partial t(\tau)}{\partial\tau} \d\tau$.   
The new action is identical to Eq.~(\ref{eq:SGR1+1}) with the lapse substitution $\eta \to \dot t$,
\beq
S[t,q] = \int \dot t\d\tau \left[ \frac{\dot q^2}{2\dot t^2} - V(q) \right].
\eeq
We will vary $S$ with respect to $t(\tau)$ ---not its endpoints of course --- even though the action is certainly independent of it (Diff invariance).
The momentum conjugate to $q$ is $p = \dot q/\dot t$, and when computing the momentum conjugate to $t$ we
again find a primary constraint
\beq\label{eq:Gparam1+1}
p_t + \frac{p^2}{2} + V(q)\approx 0,
\eeq
which is only associated with the Diff gauge symmetry (i.e., arbitrariness of $\dot t$). 
The canonical Hamiltonian is
\beq
H = \dot t \left(p_t + \frac{p^2}{2} + V(q) \right),
\eeq
which vanishes when the primary constraint holds.  Because 
\beq
\big\{p_t + \frac{p^2}{2} + V(q), H\big\}_{\rm P.B.} = 0,
\eeq
the primary constraint is preserved under $\tau$-evolution, so no secondary constraint arises.

Canonical quantization imposes the operator version of the primary constraint Eq.~(\ref{eq:Gparam1+1}),
\beq
\i\frac{\partial}{\partial t}\psi(q,t,\tau) = \left(\frac{\hat p^2}{2} + V(q)\right)\psi(q,t,\tau).
\eeq
This implies $\hat H\psi = 0$ and so $\psi(q,t,\tau) = \psi(q,t)$.  We have arrived at the familiar non-relativistic Schr\"odinger equation, and no problem of time exists.  
We could say that the ``degree of freedom" $t$ plays the role of the clock, and it is able to do so because its conjugate
momentum $p_t$ appears linearly in the constraint Eq.~(\ref{eq:Gparam1+1}).  

As expected, diffeomorphism invariance does not cause the problem of time.  Only when the lapse function was a Lagrange multiplier did a secondary constraint arise.  
The derivative appearing in the lapse function $\dot t$ creates a natural physical foliation, defined by constant $t$ surfaces.
This hints at a strategy, used here as well as in references  \cite{Kuchar:1990vy, Brown:1994py, Giesel:2007wn, Husain:2011tk}, to evade the problem, namely the use of a Lagrange multiplier to constrain the lapse function to be a time derivative.  When the foliation is no longer totally arbitrary, it can appear in the Wheeler--DeWitt equation and generate physical time evolution.

%\subsec{The correspondence principle}
Quantum mechanics is more fundamental than classical mechanics in the sense that it contains classical mechanics as a limit, but the reverse is not true.    Each individual classical trajectory can be reproduced as an $\hbar \to 0$  limit of a minimum-uncertainty quantum state.  
For example, a classical non-relativistic point particle trajectory $q_{\rm cl}(t)$ with $t \in [t_{\rm i},t_{\rm f}]$ is in correspondence with the distribution
limit of the wavefunction squared:
\beq
q_{\rm cl}(t) \quad \longleftrightarrow \quad \lim_{\hbar\to0}|\Psi_{\rm cl}(q,t)|^2 = \delta(q_{\rm cl}(t) - q) ,
\eeq
where $\Psi_{\rm cl}$ is a wavepacket of minimum time-averaged position uncertainty.
Because quantum mechanics is more fundamental, we should expect that some states (e.g., those with small quantum numbers) do not have classical interpretations, whereas all classical states should arise in the $\hbar \to 0$ limit of some quantum theory.

The story becomes more complicated in the presence of gauge symmetries.  At the classical level, gauge symmetries correspond to the appearance of arbitrary functions of time, i.e., the gauge choice \cite{Dirac:1967,Henneaux:1992ig}.  In the language of Dirac's mechanics, first-class primary constraints always correspond to gauge symmetries.  Passage to quantum mechanics then requires that the wavefunction be annihilated by the gauge generators, because a quantum gauge symmetry means that the physical Hilbert space is orthogonal to all gauge generators. 

Because the foliation is not determined by the equations of motion, it is a gauge choice.\footnote{It is only unphysical if {\em spacetime} is the only observable.} This leads to the problem of time in canonical quantum gravity, where the wavefunctional $\Psi[\bar g_{ij},t)$ for the spatial geometry $\bar g_{ij}(x)$, does not evolve in time, even though the classical configurations  $(\bar g_{ij}(x,t),\Sigma_t)$ do evolve in (an arbitrarily chosen but coordinate invariant) time.   Here $\Sigma_t$ is a foliation by space-like hypersurfaces.  
%\subsec{The model}
Much  like unimodular gravity \cite{Anderson:1971pn}, we will modify the Einstein-Hilbert action by constraining the metric volume element using a Lagrange multiplier.  However, we do not use a non-dynamical background volume form.  Instead we use a dynamical scalar field $\chi$ whose velocity will provide a local clock.  We propose the full gravitational action (suppressing boundary terms) is
\beq
S[g,\Lambda,\chi] = \int \hspace{-2mm}\sqrt{-|g|}\d^4x \left[R - 2 \Lambda(1 - \nabla^2\chi)\right]. %+ S_{\rm matter}.
\eeq

The equations of motion are
\beq
\nabla^2 \chi &=& 1,\label{eq:foliatoneom}\\
\nabla^2\Lambda &=& 0,\label{eq:Lambdaeom}\\
\hspace{-12mm}R_{\mu\nu}-\thf R g_{\mu\nu} + (\Lambda +\partial_\rho\Lambda\partial^\rho\chi) g_{\mu\nu}\nonumber \\ -\partial_{\mu}\Lambda\partial_{\nu}\chi -\partial_{\nu}\Lambda\partial_{\mu}\chi&=& 0.\label{eq:einsteineom}
\eeq
The equation of motion for $\chi$ coincides with that of a scalar field with an exactly linear potential.
The global symmetry corresponding to $\chi$-translation invariance gives rise to a conserved current
$j_\mu = \partial_\mu\Lambda.$ 
 
If we choose constant $\Lambda$, the $\chi$-field does not contribute to the Einstein equations, much like the harmonic coordinate model \cite{Kuchar:1991pq}.  Thus this theory contains as a subset all solutions of general relativity for any value of cosmological constant.  More generally, solutions resemble quintom \cite{Feng:2004ad} dark energy.   Although the bilinear kinetic term would appear to necessarily suffer from a ghost instability, this is not the case, because the Hamiltonian constraint\footnote{Similarly, bosonic string theory has no world-sheet ghost instability, despite the presence of a negative kinetic term.} ensures $\Lambda$ is not an independent field  \cite{Kuchar:1991pq}.

%

%\subsec{ADM decomposition}
The identity of the gravitational degrees of freedom can be elucidated by writing the metric in the ADM form,
\beq
\d s^2 = -N^2\d t^2 + \bar g_{ij}(\d x^i + N^i \d t)(\d x^j + N^j \d t),
\eeq
because then the lapse function $N$ and shift vector $N^i$ appear without time derivatives in the action and so lead to primary constraints $\pi_N \approx 0$, $\pi_{N^i} \approx 0$.
The Hamiltonian density of our theory is 
\beq
{\cal H} = N {\cal H}_0 + N^i{\cal H}_i,
\eeq
where
\begin{align}
{\cal H}_0 = \sqrt{|\bar g|}\Big[&\frac{\pi_{ij}\pi^{ij}}{|\bar g|} - \frac{({\pi^i}_i)^2}{2|\bar g|} - \bar{R}\nonumber \\&+ \frac{\pi_\Lambda \pi_\chi}{2|\bar g|}+2\partial_i\Lambda\partial^i\chi+2\Lambda \Big],\\  
{\cal H}_i =  -2 \bar\nabla_j &{\pi^j}_i + \pi_\Lambda \partial_i\Lambda + \pi_\chi \partial_i\chi,
\end{align}
where $\bar{R}$ and $\bar \nabla_i$ are computed from the spatial metric $\bar g_{ij}$, and $\pi^{ij}$ is conjugate to it.
The primary constraints are preserved under time evolution when
the Hamiltonian and momentum (secondary) constraints are satisfied,
\begin{align}
0 = \frac{\partial{\cal H}}{\partial N} ={\cal H}_0, 
\qquad 
0 = \frac{\partial{\cal H}}{\partial N^i} = {\cal H}_i.
\end{align}

We can use the momentum constraints to eliminate $\partial_i\Lambda$ from the Hamiltonian constraint,
which, upon canonical quantization ($\Lambda(x) \to \i\delta/\delta \pi_\Lambda(x)$), becomes the Tomonaga--Schwinger equation (a local version of the Schr\"odinger equation),
\beq
-2\i\frac{\delta}{\delta \pi_\Lambda(x)}\Psi[\pi_\Lambda,\chi,\bar g] = \hat h_\Lambda(x) \Psi[\pi_\Lambda,\chi,\bar g] ,
\eeq
with 
\begin{align}
\hat h_\Lambda(x) = & \; \frac{\hat\pi_{ij}\hat\pi^{ij}}{|\bar g|} - \frac{({\hat\pi^i}_{\;\,i})^2}{2|\bar g|} - \bar{R} + \frac{\pi_\Lambda \hat\pi_\chi}{2|\bar g|}       \nonumber  \\
& -\frac{2\hat\pi_\chi}{\pi_\Lambda}\partial_i\chi\partial^i\chi   +   \frac{4}{\pi_\Lambda}\partial_i\chi \bar\nabla_j\hat\pi^{ij} ,  \\
\hat\pi^{ij} =& -\i\frac{\delta}{\delta \bar g_{ij}(x)} , \\
\hat\pi_\chi =& -\i\frac{\delta}{\delta \chi(x)}.
\end{align}
To turn this into an ordinary functional Schr\"odinger equation, we need to choose a foliation $\pi_\Lambda^t(x)$.  This is a choice for the value of $\pi_\Lambda$ at each spatial point $x$ that is monotonic in the parameter $t$.  Notice that like all conjugate momenta here, $\pi_\Lambda$ is a tensor {\em density}, and so makes an unusual time parameter \cite{Isham:1992ms}.  We cannot use $\sqrt{|\bar g|}$ to factor out the volume form, because the spatial metric is an operator, not a c-number.  The natural choice is just the product of $t$ and a fiducial spatial volume element $\bar\mu$:
\beq
\pi_\Lambda^t(x) = -2 t\bar\mu.
\eeq
Classically, this corresponds to a lapse function
\beq
N = \frac{\bar\mu}{\sqrt{|\bar g|}(1-\bar\nabla^2\chi)}.
\eeq
It is far from clear that different choices of foliation are equivalent \cite{Torre:1998eq}, but because there is a natural choice, $\bar\mu = 1$, this may not matter.  Then, because
\beq
\frac{\partial}{\partial t} \Psi[\pi_\Lambda^t]  =  \left.\int \d^3x \frac{\partial\pi_\Lambda^t(x)}{\partial t}\frac{\delta}{\delta \pi_\Lambda(x)} \Psi[\pi_\Lambda]\right|_{\pi_\Lambda = \pi_\Lambda^t},
\eeq
we can write the time dependent functional Schr\"odinger equation for gravity
\beq
\i\frac{\partial}{\partial t}\Psi(t,\bar g,\chi] = \left. \int \d^3x\, \bar\mu\hat h_\Lambda(x)\right|_{\pi_\Lambda = -2 t\bar\mu}\hspace{-1mm}\Psi(t,\bar g,\chi].
\eeq

In $d=4$ spacetime dimensions the spatial metric has $d(d-1)/2 = 6$ components, but the $d-1 = 3$ momentum constraints reduce this to $(d-1)(d-2)/2 = 3$ degrees of freedom, the correct number for gravity plus a scalar field $\Lambda$.  Hence, the dynamical content is simply the spatial geometry, as well as the $\chi$ field.  Note that  $\pi_\chi \geq 0$, (i.e., $\dot \Lambda \geq 0$) in order for the gradient energy in the $\chi$-field to be nonnegative.  Given appropriate initial conditions for $\Lambda$, this ghost-free condition can be maintained for positive times, thanks to the conservation of the current $\partial_\mu \Lambda$.  Note that constraining $\pi_\chi \geq 0$ does not restrict the geometrical phase space.  
%This avoids a shortcoming of the incoherent dust model \cite{Kuchar:1990vy}.

We have neglected to resolve operator ordering ambiguities, nor have we attempted to regularize UV divergences, since this task is likely impossible given the nonrenormalizability of gravity in 3+1 and higher dimensions \cite{Shomer:2007vq}.  Nevertheless, this equation may prove to be a useful tool for understanding infra-red phenomena such as dark energy, eternal inflation, the measure problem, and the black-hole information paradox.

%\subsec{Mini-superspace} 
The solutions to the equations of motion are straightforward in the mini-superspace ansatz
\beq
\d s^2 &=& - N^2(t) \d t^2 + a^2(t) \d\Omega_3^2,
\eeq
and we assume $\Lambda$ and $\chi$ depend only on $t$.  We will choose the gauge $N=1$.
The scalar equations of motion are
\beq
\ddot \Lambda = -3 H \dot \Lambda,\qquad
\ddot\chi = - 3 H \dot\chi - 1.
\eeq
There is a conserved quantity associated with global $\chi$-translation invariance, namely
\beq
p_\chi =  4\pi^2 a^3 \dot\Lambda.
\eeq

Extremizing the action with respect to the lapse $N$ gives the Friedmann equation
\begin{align}
H^2 &= \frac{\Lambda}{3} - \frac{1}{a^2} + \frac{\dot\Lambda\dot\chi}{3} \\
        &= \frac{\Lambda}{3} - \frac{1}{a^2} + \frac{ p_\chi \dot\chi}{4\pi^2a^3},
\end{align}
where $H = \dot a/a$.
Notice that if we choose initial conditions $\Lambda > 0$, $\dot \Lambda = 0$, we find global de Sitter space is a solution.  

%The scalar $\chi$, which now has no backreaction on the geometry, is
%\beq
%\chi &=& \frac{-\ln \cosh \sqrt{\tfrac{\Lambda}{3}}t + {\rm sech}^2 \sqrt{\tfrac{\Lambda}{3}}t}{\Lambda},
%\eeq
%modulo periodicity.

If we perturb away from an expanding de Sitter space by choosing a positive initial $\dot\Lambda$, the solution rapidly approaches de Sitter space, because the energy density in
the homogeneous $\dot\Lambda$ perturbation decays like $1/a^3$, although it grows relative to the critical density during the radiation and matter eras, raising the possibility of
detection.  
The canonical Hamiltonian
\beq
H_{\rm c} = N \left(\frac{p_\Lambda p_\chi}{4\pi^2a^3} - \frac{p_a^2}{48\pi^2 a} - 12\pi^2 a + 4\pi^2a^3\Lambda  \right),
\eeq
vanishes according to to the single secondary constraint
\beq
0 = -\frac{p_\Lambda p_\chi}{48\pi^4a^6} + \frac{p_a^2}{576\pi^4 a^4} + \frac{1}{a^2} - \frac{\Lambda}{3}.
\eeq
The generator of time translations should appear linearly in the constraint equation, and so $\Lambda$ is the only sensible choice.  
Time is therefore played by $p_\Lambda$, obeying
\beq
p_\Lambda = 4\pi^2 a^3 \dot\chi~, \qquad \dot{p}_\Lambda = - 4\pi^2 a^3.
\eeq
Because the 4-volume increases like $\d V = 2\pi^2a^3\d t$, the clock measures 4-volume:
\beq
\frac{\d p_\Lambda}{\d V} = -2.
\eeq

%\subsec{Relation to Unimodular gravity}
This theory has some similarities with unimodular gravity \cite{Anderson:1971pn, Unruh:1988in, Unruh:1989db}, namely the arbitrariness of the cosmological constant.  Indeed, if in the action we replaced $\partial_\mu\chi$ with a (Hodge dual) three-form potential
$\Lambda(1-\nabla^2\chi) \mapsto \Lambda(1 - \epsilon^{\mu\nu\rho\sigma}\partial_\mu A_{\nu\rho\sigma})$,
 we would precisely recover the background independent form of unimodular gravity \cite{Henneaux:1989zc}.  However, because the dynamical $\Lambda$ is then constrained to be constant, the problem of time remains \cite{Kuchar:1991xd}, since the wavefunctional $\Psi[\bar g_{ij},t)$ would be invariant under all 4-volume preserving deformations of the Cauchy surface $\Sigma_t$.  Thus only one of the infinitely-many fingers of time is successfully ungauged by unimodular gravity. 

%\subsec{Discussion}
The model we have presented is unusual for a theory of gravity in that it has a natural arrow of time and foliation, although this foliation need not be imprinted on the Einstein tensor.  Because the lapse function is inversely proportional to the spatial volume element, expansion is not rewarded in the sense that the total four-volume of the universe is not dominated by the region with the largest expansion rate.
This fact allows conditional probabilities of what observers should measure to be compatible with our observation of an old universe with a small cosmological constant, evading what is known as the youngness paradox \cite{Guth:2007ng} .

\acknowledgments
We thank Jose Blanco-Pillado, Larry Ford, Jaume Garriga, Alan Guth, Ali Masoumi, Ken Olum, and Alex Vilenkin for helpful discussions.  Funding was provided through NSF grant PHY-1213888.

\end{document}